\documentclass[]{article}

\usepackage[utf8]{inputenc}
\usepackage{amsmath, amssymb, amsfonts}
\usepackage{amsthm}
\usepackage{bm}
\usepackage{graphicx}
\usepackage{fancyhdr}
\usepackage{titlesec}
\usepackage{indentfirst}
\usepackage{booktabs}
\usepackage{multirow}
\usepackage{verbatim}
\usepackage{xcolor}
\usepackage{subcaption}

\usepackage[page,header]{appendix}
\usepackage{titletoc}
\usepackage{relsize}
\usepackage[colorlinks=true, linkcolor=blue, citecolor=blue, urlcolor=blue]{hyperref}

\allowdisplaybreaks 

\newcommand{\rd}{\,\mathrm{d}}

\newcommand{\bv}{\mathbf{v}}
\newcommand{\bx}{\mathbf{x}}

\newcommand{\bB}{\mathbf{B}}
\newcommand{\bE}{\mathbf{E}}
\newcommand{\bF}{\mathbf{F}}

\newcommand{\bJ}{\mathbf{J}}

\newcommand{\ctrl}{\mathrm{ctrl}}

\numberwithin{equation}{section}
\newtheorem{theorem}{Theorem}[section]

\newtheorem{remark}[theorem]{Remark}

\begin{document}
	
	\title{Stabilization of kinetic plasma instabilities via \\ 
    instantaneous external field control}
	
	\author{Jiyoung Yoo\footnote{Department of Applied Mathematics, University of Washington, Seattle, WA 98195, USA (\texttt{jyyoo496@uw.edu}).}, \ 
		Jingwei Hu\footnote{Department of Applied Mathematics, University of Washington, Seattle, WA 98195, USA (\texttt{hujw@uw.edu}). Corresponding author. } }
	
	\maketitle
	
\begin{abstract}
We propose an instantaneous control framework for the stabilization of kinetic plasma instabilities using external fields. The control problem is formulated as a PDE-constrained optimization that accounts for nonlinear plasma dynamics, while the control is determined at each time step to avoid global-in-time optimization. The Discretize-then-Optimize (DtO) approach is adopted, with the forward equations first discretized using a particle-in-cell (PIC) method. The subsequent control computation is derived from the discrete particle update, ensuring consistency with the forward solver. The control objective is formulated in terms of electromagnetic field energy and low-order moments rather than the full distribution function, making the optimization more amenable to practical implementation. The framework is applied to both electrostatic and electromagnetic settings and evaluated on several benchmark problems, including the two-stream, bump-on-tail, Dory--Guest--Harris, and Weibel instabilities. Across these tests, the control successfully suppresses field energy growth and keeps the distribution function close to its initial state. These results demonstrate the effectiveness of the proposed strategy in suppressing kinetic plasma instabilities.
\end{abstract}





\begin{flushleft}

\end{flushleft}

\section{Introduction}

Plasma is the most abundant state of visible matter in the universe, and its study is important across a wide range of scientific and engineering fields, including space science and fusion energy research. In fusion applications, achieving and sustaining high-temperature plasmas under stable confinement is a central challenge. Plasma instabilities pose a major obstacle to this goal, motivating extensive research into their control and stabilization.

Kinetic models describe plasma dynamics through the evolution of particle distribution functions. A widely used kinetic model for a collisionless plasma is the Vlasov equation, written here in dimensionless form as
\begin{equation}\label{eq: valsov}
    \partial_t f + \bv \cdot \nabla_{\bx} f + q \left(\bE + \bv \times \bB\right) \cdot \nabla_{\bv} f = 0,
    \quad \bx \in \Omega_{\bx} \subset \mathbb{R}^d, \ \bv \in \mathbb{R}^d,
\end{equation}
where $f=f(t,\bx,\bv)$ denotes the distribution function of a single plasma species, $q$ is the normalized species charge, and $\bE=\bE(t,\bx)$ and $\bB=\bB(t, \bx)$ denote the electric and magnetic fields, respectively.

The evolution of the distribution function $f$ in the Vlasov equation~\eqref{eq: valsov} is coupled with the evolution of the electromagnetic fields $\bE$ and $\bB$. In the electromagnetic case, the self-consistent $\bE$ and $\bB$ are obtained by solving Maxwell's equations:
\begin{align}
    \nabla_{\bx} \cdot \bE &= \rho - \rho_i, \label{Gauss's law}\\
    \nabla_{\bx} \cdot \bB &= 0, \label{Gauss's law for Mag}\\
    \partial_t \bE &= \nabla_{\bx} \times \bB - \bJ, \label{Ampere-Maxwell law}\\
    \partial_t \bB &= - \nabla_{\bx} \times \bE. \label{Faraday's law}
\end{align}
Here, $\rho=\rho(t,\bx)$ and $\bJ=\bJ(t,\bx)$ are the charge density and current density, defined by
\begin{equation}\label{eq:charge-current}
    \rho(t,\bx) = q\int_{\mathbb{R}^d} f(t,\bx,\bv)\rd{\bv}, \qquad
    \bJ(t,\bx) = q\int_{\mathbb{R}^d} f(t,\bx,\bv)\bv\rd{\bv}.
\end{equation}
The quantity $\rho_i$ in~\eqref{Gauss's law} denotes a fixed background density chosen to ensure global neutrality:
\[
\rho_i = \frac{1}{|\Omega_{\bx}|}\int_{\Omega_{\bx}} \rho \rd{\bx},
\]
where $|\Omega_{\bx}|$ is the volume of $\Omega_{\bx}$.

In the electrostatic case, the magnetic field is given externally $\bB(t, \bx) = \bB^{\mathrm{ext}}(t, \bx)$ and only the electric field is determined self-consistently. Introducing the electric potential $\phi$ such that $\bE=-\nabla_{\bx}\phi$, Gauss's law reduces to the Poisson equation
\begin{equation}\label{eqn: poisson eqn}
    -\Delta_{\bx}\phi = \rho - \rho_i.
\end{equation}

Kinetic plasma instabilities have been studied for decades \cite{Penrose60,Weibel59,DGH65}, while their control has received increasing attention in recent years. In \cite{ELMY25,CJLY26}, a pole-elimination approach based on linear stability analysis was used to design external electric fields that remove unstable roots of the dispersion relation. The method linearizes the Vlasov--Poisson system in \eqref{eq: valsov} and \eqref{eqn: poisson eqn} around an equilibrium and analyzes the corresponding dispersion relation. PDE-constrained optimization has also been explored for instability suppression using external electric fields \cite{ELWY24,LWC2508}. In these approaches, the plasma dynamics are incorporated as constraints in the optimization problem, either through the full Vlasov--Poisson system or through reduced moment systems. In the objective function, either the distribution function or moments are used to measure deviations from equilibrium. A dynamic feedback strategy was also developed in \cite{LWC2509}, where an operator maps the current state perturbation to an external electric field. In \cite{CELSY26}, control through external laser fields was investigated within the Vlasov--Maxwell system in \eqref{eq: valsov} and \eqref{Gauss's law}--\eqref{Faraday's law}. Linear stability analysis was used for transverse instabilities, while PDE-constrained optimization was employed for longitudinal instabilities when linear control is ineffective. Beyond instability suppression, optimization-based control has also been studied for plasma confinement. Instantaneous strategies have been developed to steer the plasma away from device boundaries through external magnetic fields \cite{ADFP25,ADFP26,F25}. A separate line of work on plasma confinement using magnetic fields was developed in \cite{K18,BKSW24}, where \cite{K18} established a theoretical framework for bringing the plasma distribution close to a target distribution at the final time; building on this work, \cite{BKSW24} developed a particle-in-cell (PIC)-based numerical framework for confinement around a desired phase-space region.

Existing control approaches have several limitations. Methods based on linear stability analysis have a clear theoretical foundation and provide an analytically motivated control design, but their applicability may become limited as the system evolves into a nonlinear regime. PDE-constrained optimization, on the other hand, can directly account for the nonlinear dynamics, but typically requires solving a large-scale optimization problem over the entire time horizon, which can be computationally expensive. In particular, gradient-based optimization methods often require solving an adjoint problem in addition to the forward problem, further increasing the computational cost. To alleviate these limitations, we employ an instantaneous optimization approach that determines the control field at each time step from the current plasma state. This avoids a global-in-time optimization while still accounting directly for the nonlinear dynamics. Within this framework, we consider both the Vlasov--Poisson and Vlasov--Maxwell systems, with external electric or magnetic fields used as control variables depending on the problem setting. 
Following the Discretize-then-Optimize (DtO) approach, we formulate the optimization problem directly from the fully discrete particle system. This is particularly useful for maintaining consistency between the control computation and the numerical state evolution, since the optimization is based on the same discrete dynamics used in the forward simulation. By contrast, a common approach in continuum formulations with grid-based forward solvers is to derive the optimality system at the continuous level and subsequently discretize it. In this case, the discretized optimality system may not be fully consistent with the discrete dynamics used for the forward state evolution. In our implementation, the same particle representation and time discretization are used both to determine the control and to evolve the controlled state. This also differs from the explicit instantaneous control derivations in \cite{ADFP25,ADFP26}, where a simplified time discretization is used for the control derivation and a different scheme is used for the state evolution.
The resulting control framework can then be adapted to different kinetic instabilities by choosing the target moments and control parameters accordingly. In this work, we apply it to the two-stream, bump-on-tail, and Dory--Guest--Harris instabilities in electrostatic settings, and to the Weibel instability in an electromagnetic setting.

The remainder of this paper is organized as follows. Section~\ref{sec:formulation of control} introduces our proposed control framework. In particular, Section~\ref{sec:instantaneous-pde} formulates the PDE-constrained optimization problem used to determine the instantaneous control. Section~\ref{sec:fully-discrete-system} first presents the full discretization procedures for the electrostatic and electromagnetic models, followed by the derivation of the minimizer of the resulting discrete optimization problem in Section~\ref{sec: derive-control}. Section~\ref{Sec:numerical} presents numerical experiments demonstrating the performance of the proposed control strategy. Finally, conclusions are given in Section~\ref{Sec:conclusion}.

\section{Formulation of the instantaneous control framework}
\label{sec:formulation of control}

Our goal is to suppress kinetic instabilities and keep the plasma distribution close to a desired equilibrium as it evolves according to the Vlasov--Poisson or Vlasov--Maxwell systems. Such instabilities can arise when small perturbations to an unstable equilibrium grow in time. In the electrostatic case, for example, a perturbation may produce a spatially inhomogeneous charge density and consequently a nontrivial electric field. Through the Vlasov dynamics, this field affects the particle motion, which can further amplify the perturbation and drive the system away from the equilibrium. The development of an electrostatic instability is therefore often accompanied by the growth of the electric field energy. Similarly, in the electromagnetic case, where the electric and magnetic fields are coupled through Maxwell's equations, the growth of an instability may be reflected in the electromagnetic field energy. For this reason, field energy has been used as an indicator of instability growth \cite{E24,CELSY26}. This motivates us to use the field energy as a primary quantity in the control objective. This choice also avoids directly measuring the distribution function $f$ itself. A commonly considered approach is to measure the difference between the distribution function and a desired target distribution, for example through a quantity such as $\left\|f-f_{\mathrm{target}}\right\|$ \cite{K18,ELWY24,LWC2509}. However, directly measuring the distribution function can be difficult in practice. In contrast, the electric and magnetic fields are more accessible. We therefore formulate the stabilization problem mainly in terms of reducing the field energy.

However, minimizing the field energy alone does not necessarily keep the distribution close to the desired equilibrium. Therefore, we also include moment-based quantities in the objective, which provide additional information about the distribution without measuring $f$ itself. In addition, a regularization term is added to penalize large control fields. Based on these considerations, we construct the objective functional in the general form
\begin{align}\label{def: full obj}
\mathcal{J}(f,\bF)=\mathcal{J}_{\mathrm{field}}(f,\bF)+\mathcal{J}_{\mathrm{moment}}(f,\bF)+\mathcal{J}_{\mathrm{reg}}(\bF),
\end{align}
where $\bF$ denotes the external electric or magnetic control field. The objective is minimized subject to the controlled plasma dynamics, consisting of the Vlasov equation together with the Poisson equation~\eqref{eqn: poisson eqn} or Maxwell's equations~\eqref{Gauss's law}--\eqref{Faraday's law}. 

Further details of this PDE-constrained formulation are provided in Section~\ref{sec:instantaneous-pde}. In particular, we introduce its instantaneous formulation and specify the three objective terms $\mathcal{J}_{\mathrm{field}}, \ \mathcal{J}_{\mathrm{moment}},$ and $\mathcal{J}_{\mathrm{reg}}$ in the corresponding one-step setting. 
We then describe the fully discrete formulation of the control problem in Section~\ref{sec:fully-discrete-system}.


\subsection{Instantaneous PDE-constrained optimization}\label{sec:instantaneous-pde}

To stabilize plasma instabilities, we use an external electric or magnetic field as the control variable. To show how both types of control enter the plasma dynamics, we write the controlled Vlasov equation~\eqref{eq: valsov} in the general form
\begin{align}\label{eq: controlled vlasov}
    \partial_t f
    + \bv \cdot \nabla_{\bx} f + q\left(\bE + \bE^{\ctrl}
    + \bv \times \left(\bB+\bB^{\ctrl}\right)\right)\cdot\nabla_{\bv} f = 0,
\end{align}
where $\bE^{\ctrl}=\bE^{\ctrl}(t,\bx)$ and $\bB^{\ctrl}=\bB^{\ctrl}(t,\bx)$ denote the external electric and magnetic control fields. Although both fields are included in this general form, we use either $\bE^{\ctrl}$ or $\bB^{\ctrl}$ depending on the model. The fields $\bE$ and $\bB$ are determined self-consistently or given externally.

For the general optimization formulation, we use $\bF$ here to denote either $\bE^{\ctrl}$ or $\bB^{\ctrl}$. A standard PDE-constrained control problem can then be formulated over a time interval $[0,t_{\mathrm{final}}]$ as
\begin{equation}
\begin{aligned}
\min_{\bF}\quad & \int_0^{t_{\mathrm{final}}} \mathcal{J}(f(t),\bF(t))\,\rd t \\
\text{s.t.}\quad & \text{controlled Vlasov equation~\eqref{eq: controlled vlasov} together with}\\
& \text{Poisson equation~\eqref{eqn: poisson eqn} or Maxwell's equations~\eqref{Gauss's law}--\eqref{Faraday's law}},
\end{aligned}
\end{equation}
where $\mathcal{J}$ is given by~\eqref{def: full obj}. In this formulation, the control is determined over the entire time interval $[0,t_{\mathrm{final}}]$, requiring the state evolution to be treated globally in time. To reduce the computational burden, we instead determine the control locally at each time step through an instantaneous optimization problem. Given the state at time $t^n$, the control $\bF^n$ for the evolution from $t^n$ to $t^{n+1}$ is determined by
\begin{equation}
\begin{aligned}\label{def: instan opt prob}
\bF^n=\arg\min_{\bF}\quad & \mathcal{J}^n(f,\bF) \\
\text{s.t.}\quad & \text{time-discretized version of the controlled system}\\
& \text{\eqref{eq: controlled vlasov} together with \eqref{eqn: poisson eqn} or \eqref{Gauss's law}--\eqref{Faraday's law}}.
\end{aligned}
\end{equation}
This is the instantaneous PDE-constrained optimization problem that we solve in this work. Solving this problem gives the control $\bF^n$, and the control is then applied to advance the state from $t^n$ to $t^{n+1}$. The resulting state at $t^{n+1}$ is used to determine the next control $\bF^{n+1}$, and this procedure is repeated until the final time is reached. 

We now describe the three contributions to the objective in more detail in the instantaneous setting. First, the field energy term $\mathcal{J}^n_{\mathrm{field}}$ is designed to reduce the field energy at the next time step. Thus, the control $\bF^n$ is determined based on its effect on the fields at $t^{n+1}$. In the electrostatic case, we minimize the electric field energy, while in the electromagnetic case both the electric and magnetic field energies are considered. Accordingly, the field energy contribution is given by
\begin{equation}\label{obj: field}
\mathcal{J}_{\mathrm{field}}^n(f,\bF)=
\begin{cases}
\dfrac{\gamma_{\bE}}{2}\left|\bE^{n+1}(\bF)\right|^2, & \text{electrostatic case}, \\[2mm]
\dfrac{\gamma_{\bE}}{2}\left|\bE^{n+1}(\bF)\right|^2+\dfrac{\gamma_{\bB}}{2}\left|\bB^{n+1}(\bF)\right|^2, & \text{electromagnetic case}.
\end{cases}
\end{equation}
Next, we introduce a moment penalty term $\mathcal{J}^n_{\mathrm{moment}}$ to discourage deviations of the low-order moments from their prescribed equilibrium values. This plays an important role because, when only the field energy is minimized, the control is selected solely to reduce the field energy and may drive the distribution far from the desired equilibrium. The moment-based term therefore complements the field energy, which provides only an indirect measure of the distribution. As in the field energy term, the moments are evaluated at $t^{n+1}$ based on the state resulting from the current control. 
We use the charge density $\rho$ and current density $\bJ$ defined in~\eqref{eq:charge-current}, together with the second-order velocity moment
\begin{equation}\label{def: 2nd moment}
\mathcal{E}=q\int_{\mathbb{R}^d}\bv\otimes\bv\,f\,\rd\bv.
\end{equation}
The moment contribution is then given by
\begin{equation}\label{obj: moment}
\begin{aligned}
\mathcal{J}_{\mathrm{moment}}^n(f,\bF)={}&\frac{\alpha_{\rho}}{2}\left|\rho^{n+1}(\bF)-\bar{\rho}\right|^2+\frac{\alpha_{\bJ}}{2}\left|\bJ^{n+1}(\bF)-\bar{\bJ}\right|^2+\frac{\alpha_{\mathcal{E}}}{2}\left|\mathcal{E}^{n+1}(\bF)-\bar{\mathcal{E}}\right|^2,
\end{aligned}
\end{equation}
where $\bar{\rho}$, $\bar{\bJ}$, and $\bar{\mathcal{E}}$ denote the target moments. Finally, we include a regularization term to prevent the magnitude of the control field from becoming too large, using the standard quadratic form
\begin{equation}\label{obj: regularization}
\mathcal{J}_{\mathrm{reg}}^n(\bF)=\frac{\beta_{\bF}}{2}|\bF|^2.
\end{equation}
In all of the above, $\gamma_{\bE}$, $\gamma_{\bB}$, $\alpha_{\rho}$, $\alpha_{\bJ}$, and $\alpha_{\mathcal{E}}$ are nonnegative weighting parameters, while $\beta_{\bF}>0$ is the regularization parameter.

\subsection{Fully discrete formulation of the control problem}\label{sec:fully-discrete-system}

In this section, we describe the spatial and temporal discretization of the instantaneous optimization problem defined in~\eqref{def: instan opt prob}. We use a standard particle-in-cell (PIC) discretization for the controlled system, a commonly used particle method that introduces grid cells to compute the electric and magnetic fields. For the time integration, we use the Verlet scheme in the electrostatic case and the Boris scheme~\cite{Boris1970} in the electromagnetic case. Both schemes are second-order accurate in time and are widely used in PIC simulations. 

\subsubsection{Particle-in-cell (PIC) discretization}
In the particle method, the distribution function $f(t,\bx,\bv)$ is approximated using $N$ particles:
\begin{align}\label{eq: particle solution}
    f^N(t,\bx,\bv)
    = \sum_{p=1}^N w_p
    \delta\left(\bx-\bx_p(t)\right)
    \delta\left(\bv-\bv_p(t)\right).
\end{align}
Here, $w_p$, $\bx_p(t)$, and $\bv_p(t)$ denote the weight, position, and velocity of particle $p$, respectively. The particle weights remain constant in time. The particle positions $\bx_p$ and velocities $\bv_p$ are evolved along the characteristics of the controlled Vlasov equation~\eqref{eq: controlled vlasov}:
\begin{align}
\begin{aligned}\label{eqn: controlled particle method}
\frac{\rd \bx_p}{\rd t} &= \bv_p, \\
\frac{\rd \bv_p}{\rd t} &= q\left(\bE(t, \bx_p)+\bE^{\ctrl}(t, \bx_p)+\bv_p\times\left(\bB(t, \bx_p)+\bB^{\ctrl}(t, \bx_p)\right)\right).
\end{aligned}
\end{align}

To couple the particles with the spatial grid in the PIC discretization, we introduce the spatial shape function $S_\eta(\cdot)$, which is a nonnegative, symmetric, normalized kernel with width $\eta$. Using the particle representation~\eqref{eq: particle solution}, the charge density defined in~\eqref{eq:charge-current} can be approximated as
\begin{align*}
\rho(t,\bx)
&\approx q\int_{\mathbb{R}^d}f^N(t,\bx,\bv)\,\rd\bv=q\sum_{p=1}^N w_p\delta\left(\bx-\bx_p(t)\right)
\approx q\sum_{p=1}^N w_p S_\eta\left(\bx-\bx_p(t)\right).
\end{align*}
Applying the same procedure to the current density, we obtain the following approximations:
\begin{align}
\rho(t,\bx_h) &\approx \rho_h(t):=q\sum_{p=1}^N w_p S_\eta\left(\bx_h-\bx_p(t)\right), \label{def: rho_h}\\
\bJ(t,\bx_h) &\approx \bJ_h(t):=q\sum_{p=1}^N w_p\bv_p(t)S_\eta\left(\bx_h-\bx_p(t)\right). \label{def: J_h}
\end{align}
Here, $\bx_h$ denotes the spatial grid points, and $\rho_h$ and $\bJ_h$ denote the approximations of the 0th- and 1st-order moments on the grid. This particle-grid coupling allows the self-consistent electric and magnetic fields to be computed on the spatial grid through the Poisson or Maxwell equations using $\rho_h$ and $\bJ_h$. The resulting fields are then interpolated back to the particle positions, since the particle update in \eqref{eqn: controlled particle method} requires the field values at the particle locations. We use the same spatial shape function $S_\eta$ for both the regularization and interpolation steps, and hence
\begin{align*}
\bE\left(t,\bx_p(t)\right) &\approx \bE_p(t):=\eta^d\sum_h \bE_h(t)S_\eta\left(\bx_p(t)-\bx_h\right), \\
\bB\left(t,\bx_p(t)\right) &\approx \bB_p(t):=\eta^d\sum_h \bB_h(t)S_\eta\left(\bx_p(t)-\bx_h\right).
\end{align*}

The PIC method~\eqref{eqn: controlled particle method} does not evolve the distribution function directly. Instead, it evolves the positions and velocities of the representative particles with the electromagnetic fields computed on the grid. If the phase-space distribution function is needed, it can be reconstructed by further introducing a velocity shape function $S_\epsilon(\cdot)$:
\begin{align}\label{eq: regularized particle solution}
f_{\eta,\epsilon}^N(t,\bx,\bv)=\sum_{p=1}^N w_p S_\eta\left(\bx-\bx_p(t)\right)S_\epsilon\left(\bv-\bv_p(t)\right),
\end{align}
where $S_\epsilon$ is a normalized kernel with width $\epsilon$, satisfying the same properties as $S_\eta$. However, since our objective functional does not require the distribution function itself, there is no need to reconstruct $f^N_{\eta, \epsilon}$ during the control procedure.


\subsubsection{Electrostatic case}\label{sec: electrstatic verlet}

In the electrostatic case, we use the external electric field $\bE^{\ctrl}$ as the control and set $\bB^{\ctrl}=0$. The self-consistent electric field is computed from the Poisson equation~\eqref{eqn: poisson eqn}. The electric control field is added to the intermediate self-consistent electric field $\bE_h^{*,n}$ used in the Verlet update. We then define
\begin{equation}
\bE_h^{\mathrm{tot},*,n}=\bE_h^{*,n}+\bE_h^{\ctrl,n+1/2}.
\end{equation} 
The controlled Verlet update is written as
\begin{align}
\begin{aligned}\label{controlled verlet}
\bx_p^{*,n} &= \bx_p^n + \frac{\Delta t}{2}\bv_p^n, \\
\bv_p^{n+1} &= \bv_p^n + \Delta t\,q\left(\bE_p^{\mathrm{tot},*,n}+\bv_p^{n+1/2}\times\bB^{\mathrm{ext}}\!\left(t^{n+1/2},\bx_p^{*,n}\right)\right), \\
\bx_p^{n+1} &= \bx_p^{*,n} + \frac{\Delta t}{2}\bv_p^{n+1}.
\end{aligned}
\end{align}
where $\bv_p^{n+1/2}=\left(\bv_p^n+\bv_p^{n+1}\right)/2$. Here, $\bE_h^{\ctrl,n+1/2}$ is piecewise constant over the spatial grid cells. The total field $\bE_h^{\mathrm{tot},*,n}$ is then interpolated to the particle positions $\bx_p^{*,n}$ to obtain $\bE_p^{\mathrm{tot},*,n}$.

First, to evaluate $\mathcal{J}_{\mathrm{field}}^n$, we use the latest electric field affected by the current control before the next control is applied. In the Verlet scheme, this field is obtained after advancing the particle positions by an additional half step,
\begin{equation}\label{verlet x^*,n+1}
\bx_p^{*,n+1}=\bx_p^{n+1}+\frac{\Delta t}{2}\bv_p^{n+1},
\end{equation}
and solving the Poisson equation using this particle state. We denote this electric field by $\bE_h^{*,n+1}$. It depends on the control $\bE_h^{\ctrl,n+1/2}$ through the particle evolution in~\eqref{controlled verlet} and the additional half-step update~\eqref{verlet x^*,n+1}.

Next, the moment terms in $\mathcal{J}^n_{\mathrm{moment}}$ are naturally evaluated at time $t^{n+1}$, after completing the controlled Verlet update in~\eqref{controlled verlet}. Using the particle state $(\bx_p^{n+1},\bv_p^{n+1})$, we compute
\begin{equation}
\rho_h^{n+1}, \qquad \bJ_h^{n+1}, \qquad \mathcal{E}_h^{n+1}.
\end{equation}

Since $f$ no longer appears explicitly at the fully discrete level and the state is represented through the controlled Verlet scheme~\eqref{controlled verlet}, we omit $f$ from the arguments of $\mathcal{J}^n$. Accordingly, the discrete objective functional is written only in terms of the control variable as
\begin{align}\label{obj: electrostatic}
\begin{aligned}
\mathcal{J}^n\!\left(\bE_h^{\ctrl,n+1/2}\right)
&= \frac{\gamma_{\bE}}{2}\left\|\bE_h^{*,n+1}\!\left(\bE_h^{\ctrl,n+1/2}\right)\right\|_h^2 \\
&\quad+ \frac{\alpha_{\rho}}{2}\left\|\rho_h^{n+1}\!\left(\bE_h^{\ctrl,n+1/2}\right)-\bar{\rho}\right\|_h^2
+\frac{\alpha_{\bJ}}{2}\left\|\bJ_h^{n+1}\!\left(\bE_h^{\ctrl,n+1/2}\right)-\bar{\bJ}\right\|_h^2 \\
&\quad+ \frac{\alpha_{\mathcal{E}}}{2}\left\|\mathcal{E}_h^{n+1}\!\left(\bE_h^{\ctrl,n+1/2}\right)-\bar{\mathcal{E}}\right\|_h^2
+\frac{\beta_{\bE^{\ctrl}}}{2}\left\|\bE_h^{\ctrl,n+1/2}\right\|_h^2.
\end{aligned}
\end{align}
Here, $\|\cdot\|_h$ denotes the discrete $L^2$ norm on the spatial grid, induced by the inner product
\begin{equation}\label{def: norm}
\langle \mathbf{G},\mathbf{H}\rangle_h = h^d\sum_{\mathbf{i}}\mathbf{G}_{\mathbf{i}}\cdot\mathbf{H}_{\mathbf{i}}, \qquad \|\mathbf{G}\|_h^2 = \langle\mathbf{G},\mathbf{G}\rangle_h,
\end{equation}
for grid quantities $\mathbf{G}$ and $\mathbf{H}$, where $\mathbf{i}$ is the $d$-dimensional multi-index associated with the spatial grid.

Therefore, at each time step, we solve the following constrained optimization problem:
\begin{equation}\label{opt prob verlet}
\min_{\bE_h^{\ctrl,n+1/2}}
\mathcal{J}^n\!\left(\bE_h^{\ctrl,n+1/2}\right) \text{ as given in } \eqref{obj: electrostatic},
\quad \text{s.t. } \eqref{controlled verlet}.
\end{equation}
Once the optimal control $\bE_h^{\ctrl,n+1/2}$ is determined, the particle state is advanced to the next time step using \eqref{controlled verlet}.

\subsubsection{Electromagnetic case}\label{sec: electromag boris}

In the electromagnetic case, we use the external magnetic field $\bB^{\ctrl}$ as the control and set $\bE^{\ctrl}=0$. The magnetic control field is added to the intermediate self-consistent magnetic field $\bB_h^{*,n}$ used in the Boris update. We then define
\begin{equation}
\bB_h^{\mathrm{tot},*,n}=\bB_h^{*,n}+\bB_h^{\ctrl,n+1/2}.
\end{equation}
The controlled Boris update is written as
\begin{align}
\begin{aligned}\label{controlled boris}
\bx_p^{*,n} &= \bx_p^n+\frac{\Delta t}{2}\bv_p^n, \\
\bE_h^{*,n} &= \bE_h^n+\frac{\Delta t}{2}\left(\nabla_h\times\bB_h^n-\bJ_h^n\right), \\
\bB_h^{*,n} &= \bB_h^n-\frac{\Delta t}{2}\nabla_h\times\bE_h^n, \\
\bv_p^{n+1} &= \bv_p^n+\Delta t\,q\left(\bE_p^{*,n}+\bv_p^{n+1/2}\times\bB_p^{\mathrm{tot},*,n}\right), \\
\bx_p^{n+1} &= \bx_p^{*,n}+\frac{\Delta t}{2}\bv_p^{n+1}, \\
\bE_h^{n+1} &= \bE_h^n+\Delta t\left(\nabla_h\times\bB_h^{\mathrm{tot},*,n}-\bJ_h^{*,n}\right), \\
\bB_h^{n+1} &= \bB_h^n-\Delta t\,\nabla_h\times\bE_h^{*,n},
\end{aligned}
\end{align}
where 
\begin{align}
\bJ_h^n=q\sum_{p=1}^N w_p\,\bv_p^n S_h\!\left(\bx_h-\bx_p^n\right), \qquad
\bJ_h^{*,n}=q\sum_{p=1}^N w_p\,\bv_p^{n+1/2} S_h\!\left(\bx_h-\bx_p^{*,n}\right).
\end{align}

As in the electrostatic case, $\bB_h^{\ctrl,n+1/2}$ is piecewise constant over the spatial grid cells, and $\bB_h^{\mathrm{tot},*,n}$ is interpolated to the particle positions $\bx_p^{*,n}$. The objective functional is constructed in the same form as \eqref{obj: electrostatic}. According to \eqref{obj: field}, both the electric and magnetic fields are required in the field energy contribution $\mathcal{J}_{\mathrm{field}}^n$. We evaluate them at the following time levels:
\begin{align}
\bE_h^{*,n+1} &= \bE_h^{n+1}+\frac{\Delta t}{2}\left(\nabla_h\times\bB_h^{n+1}-\bJ_h^{n+1}\right), \\
\bB_h^{*,n+1} &= \bB_h^{n+1}-\frac{\Delta t}{2}\nabla_h\times\bE_h^{n+1}.
\end{align}
The moment terms are likewise evaluated at time $t^{n+1}$, namely $\rho_h^{n+1}$, $\bJ_h^{n+1}$, and $\mathcal{E}_h^{n+1}$. Finally, we write the discrete objective functional with the control dependence shown explicitly as
\begin{align}\label{obj: electromag}
\begin{aligned}
\mathcal{J}^n\!\left(\bB_h^{\ctrl,n+1/2}\right)
&= \frac{\gamma_{\bE}}{2}\left\|\bE_h^{*,n+1}\!\left(\bB_h^{\ctrl,n+1/2}\right)\right\|_h^2
+\frac{\gamma_{\bB}}{2}\left\|\bB_h^{*,n+1}\!\left(\bB_h^{\ctrl,n+1/2}\right)\right\|_h^2 \\
&\quad+\frac{\alpha_{\rho}}{2}\left\|\rho_h^{n+1}\!\left(\bB_h^{\ctrl,n+1/2}\right)-\bar{\rho}\right\|_h^2
+\frac{\alpha_{\bJ}}{2}\left\|\bJ_h^{n+1}\!\left(\bB_h^{\ctrl,n+1/2}\right)-\bar{\bJ}\right\|_h^2 \\
&\quad+\frac{\alpha_{\mathcal{E}}}{2}\left\|\mathcal{E}_h^{n+1}\!\left(\bB_h^{\ctrl,n+1/2}\right)-\bar{\mathcal{E}}\right\|_h^2
+\frac{\beta_{\bB^{\ctrl}}}{2}\left\|\bB_h^{\ctrl,n+1/2}\right\|_h^2.
\end{aligned}
\end{align}

Therefore, at each time step, we solve the following constrained optimization problem:
\begin{equation}\label{opt prob boris}
\min_{\bB_h^{\ctrl,n+1/2}}
\mathcal{J}^n\!\left(\bB_h^{\ctrl,n+1/2}\right) \text{ as given in } \eqref{obj: electromag},
\quad
\text{s.t. }  \eqref{controlled boris}.
\end{equation}
Once the optimal control $\bB_h^{\ctrl,n+1/2}$ is determined, the particle state is advanced to the next step using 
\eqref{controlled boris}.

\section{Derivation of the instantaneous control}\label{sec: derive-control}

We now derive the minimizers of the instantaneous optimization problems~\eqref{opt prob verlet} and~\eqref{opt prob boris}. Since the derivation follows the same procedure in both cases, we first present it for the electrostatic case, using the electric control field $\bE_h^{\ctrl}$ as the optimization variable. The corresponding electromagnetic formulation is obtained analogously and summarized afterward. 


The field and moment quantities in the discrete objective depend nonlinearly on the control through the
discrete particle dynamics. An explicit expression for the minimizer is therefore difficult to obtain.
To address this difficulty, we introduce a local linear approximation of these quantities with respect to the control variable. This yields a quadratic approximation of the objective with respect to the control variable.
The resulting quadratic problem is strictly convex
and admits a unique minimizer. This avoids the sensitivity to the initial guess that can arise when solving the original nonlinear optimization problem due to the presence of different local minima.

Given the state at time $t^n$, we omit the time indices
of the control, field, and moment quantities for simplicity when constructing the quadratic approximation $\widetilde{\mathcal{J}}^n$. For the electrostatic case, let $\overline{\bE}_h^{\ctrl}$ denote the expansion point for the local linearization. We then approximate each quantity in the electrostatic objective functional~\eqref{obj: electrostatic} to first order with respect to $\bE_h^{\ctrl}$ about $\overline{\bE}_h^{\ctrl}$. In particular,
\begin{align}
\begin{aligned}\label{linearizations}
&\bE_h(\bE_h^{\ctrl}) \approx \bE_h(\overline{\bE}_h^{\ctrl})+J_{\bE}\left(\bE_h^{\ctrl}-\overline{\bE}_h^{\ctrl}\right), \quad
\rho_h(\bE_h^{\ctrl}) \approx \rho_h(\overline{\bE}_h^{\ctrl})+J_{\rho}\left(\bE_h^{\ctrl}-\overline{\bE}_h^{\ctrl}\right), \\
&\bJ_h(\bE_h^{\ctrl}) \approx \bJ_h(\overline{\bE}_h^{\ctrl})+J_{\bJ}\left(\bE_h^{\ctrl}-\overline{\bE}_h^{\ctrl}\right), \quad \ 
\mathcal{E}_h(\bE_h^{\ctrl}) \approx \mathcal{E}_h(\overline{\bE}_h^{\ctrl})+J_{\mathcal{E}}\left(\bE_h^{\ctrl}-\overline{\bE}_h^{\ctrl}\right).
\end{aligned}
\end{align}
Note that the regularization term $\beta_{\bE^{\ctrl}}\left\|\bE_h^{\ctrl}\right\|_h^2/2$ is already quadratic in the control variable and does not require linearization. Here, $J_{\bE}$, $J_{\rho}$, $J_{\bJ}$, and $J_{\mathcal{E}}$ denote the Jacobian matrices evaluated at $\overline{\bE}_h^{\ctrl}$:
\begin{align}\label{electrostatic_jacob}
J_{\bE}=\left.\frac{\partial \bE_h}{\partial \bE_h^{\ctrl}}\right|_{\overline{\bE}_h^{\ctrl}}, \qquad
J_{\rho}=\left.\frac{\partial \rho_h}{\partial \bE_h^{\ctrl}}\right|_{\overline{\bE}_h^{\ctrl}}, \qquad
J_{\bJ}=\left.\frac{\partial \bJ_h}{\partial \bE_h^{\ctrl}}\right|_{\overline{\bE}_h^{\ctrl}}, \qquad
J_{\mathcal{E}}=\left.\frac{\partial \mathcal{E}_h}{\partial \bE_h^{\ctrl}}\right|_{\overline{\bE}_h^{\ctrl}}.
\end{align}

Substituting the linearizations~\eqref{linearizations} into the objective functional~\eqref{obj: electrostatic} gives a quadratic approximation of the objective, which we denote by $\widetilde{\mathcal{J}}$. Thus,
\begin{align}
\begin{aligned}\label{def: linearized full obj}
\mathcal{J}^n(\bE_h^{\ctrl})
\approx \widetilde{\mathcal{J}}^n(\bE_h^{\ctrl})
&= \frac{\gamma_{\bE}}{2}\left\|\bE_h(\overline{\bE}_h^{\ctrl})+J_{\bE}\left(\bE_h^{\ctrl}-\overline{\bE}_h^{\ctrl}\right)\right\|_h^2 \\ & \quad +\frac{\alpha_{\rho}}{2}\left\|\rho_h(\overline{\bE}_h^{\ctrl})+J_{\rho}\left(\bE_h^{\ctrl}-\overline{\bE}_h^{\ctrl}\right)-\bar{\rho}\right\|_h^2 \\
& \quad +\frac{\alpha_{\bJ}}{2}\left\|\bJ_h(\overline{\bE}_h^{\ctrl})+J_{\bJ}\left(\bE_h^{\ctrl}-\overline{\bE}_h^{\ctrl}\right)-\bar{\bJ}\right\|_h^2 \\
& \quad +\frac{\alpha_{\mathcal{E}}}{2}\left\|\mathcal{E}_h(\overline{\bE}_h^{\ctrl})+J_{\mathcal{E}}\left(\bE_h^{\ctrl}-\overline{\bE}_h^{\ctrl}\right)-\bar{\mathcal{E}}\right\|_h^2  +\frac{\beta_{\bE^{\ctrl}}}{2}\left\|\bE_h^{\ctrl}\right\|_h^2.
\end{aligned}
\end{align}

\noindent The first-order optimality condition $\nabla_{\bE_h^{\ctrl}}\widetilde{\mathcal{J}}=0$ can then be written as the linear system
\begin{equation}\label{eq:linear system control}
A_{\bE^{\ctrl}}\bE_h^{\ctrl}=b_{\bE^{\ctrl}},
\end{equation}
where
\begin{align}\label{def: A}
A_{\bE^{\ctrl}}
&= \beta_{\bE^{\ctrl}}I+\gamma_{\bE}J_{\bE}^T J_{\bE}
+\alpha_{\rho}J_{\rho}^T J_{\rho}
+\alpha_{\bJ}J_{\bJ}^T J_{\bJ}
+\alpha_{\mathcal{E}}J_{\mathcal{E}}^T J_{\mathcal{E}},
\end{align}
and
\begin{align}\label{def: b}
\begin{aligned}
b_{\bE^{\ctrl}}
&= \gamma_{\bE}J_{\bE}^T\left(J_{\bE}\overline{\bE}_h^{\ctrl}-\bE_h(\overline{\bE}_h^{\ctrl})\right) +\alpha_{\rho}J_{\rho}^T\left(J_{\rho}\overline{\bE}_h^{\ctrl}-\rho_h(\overline{\bE}_h^{\ctrl})+\bar{\rho}\right) \\
&\quad +\alpha_{\bJ}J_{\bJ}^T\left(J_{\bJ}\overline{\bE}_h^{\ctrl}-\bJ_h(\overline{\bE}_h^{\ctrl})+\bar{\bJ}\right) +\alpha_{\mathcal{E}}J_{\mathcal{E}}^T\left(J_{\mathcal{E}}\overline{\bE}_h^{\ctrl}-\mathcal{E}_h(\overline{\bE}_h^{\ctrl})+\bar{\mathcal{E}}\right).
\end{aligned}
\end{align}
Here, $J^T$ denotes the adjoint of the corresponding Jacobian with respect to the discrete $L^2$ inner product defined in~\eqref{def: norm}. Since $\beta_{\bE^{\ctrl}}>0$ and all other weighting parameters are nonnegative, $A_{\bE^{\ctrl}}$ is positive definite. Thus, the quadratic approximation~\eqref{def: linearized full obj} is strictly convex. All terms in $A_{\bE^{\ctrl}}$~\eqref{def: A} and $b_{\bE^{\ctrl}}$~\eqref{def: b} are known from the state at time $t^n$ and the chosen expansion point $\overline{\bE}_h^{\ctrl}$. Thus, solving the linear system~\eqref{eq:linear system control} gives the minimizer $\bE_h^{\ctrl}$ of the quadratic approximation of the objective, which serves as the instantaneous electric control.

The electromagnetic case follows the same procedure, with $\bB_h^{\ctrl}$ as the optimization variable and the objective functional given by~\eqref{obj: electromag}. Let $\overline{\bB}_h^{\ctrl}$ denote the expansion point for the local linearization. As above, we omit the time indices for simplicity. The quantities $\bE_h$, $\bB_h$, $\rho_h$, $\bJ_h$, and $\mathcal{E}_h$ are linearized with respect to $\bB_h^{\ctrl}$ about $\overline{\bB}_h^{\ctrl}$. The corresponding Jacobian matrices are
\begin{align}
\begin{aligned}\label{electromag_jacob}
&J_{\bE}=\left.\frac{\partial \bE_h}{\partial \bB_h^{\ctrl}}\right|_{\overline{\bB}_h^{\ctrl}}, \qquad
J_{\bB}=\left.\frac{\partial \bB_h}{\partial \bB_h^{\ctrl}}\right|_{\overline{\bB}_h^{\ctrl}}, \\
&J_{\rho}=\left.\frac{\partial \rho_h}{\partial \bB_h^{\ctrl}}\right|_{\overline{\bB}_h^{\ctrl}}, \qquad
J_{\bJ}=\left.\frac{\partial \bJ_h}{\partial \bB_h^{\ctrl}}\right|_{\overline{\bB}_h^{\ctrl}}, \qquad
J_{\mathcal{E}}=\left.\frac{\partial \mathcal{E}_h}{\partial \bB_h^{\ctrl}}\right|_{\overline{\bB}_h^{\ctrl}}.
\end{aligned}
\end{align}
The first-order optimality condition for the resulting quadratic approximation gives
\begin{equation}\label{eq:linear system control em}
A_{\bB^{\ctrl}}\bB_h^{\ctrl}=b_{\bB^{\ctrl}},
\end{equation}
where
\begin{align}\label{def: A em}
A_{\bB^{\ctrl}}=\beta_{\bB^{\ctrl}}I+\gamma_{\bE}J_{\bE}^T J_{\bE}+\gamma_{\bB}J_{\bB}^T J_{\bB}+\alpha_{\rho}J_{\rho}^T J_{\rho}+\alpha_{\bJ}J_{\bJ}^T J_{\bJ}+\alpha_{\mathcal{E}}J_{\mathcal{E}}^T J_{\mathcal{E}},
\end{align}
and
\begin{align}\label{def: b em}
\begin{aligned}
b_{\bB^{\ctrl}}
&=\gamma_{\bE}J_{\bE}^T\left(J_{\bE}\overline{\bB}_h^{\ctrl}-\bE_h(\overline{\bB}_h^{\ctrl})\right)+\gamma_{\bB}J_{\bB}^T\left(J_{\bB}\overline{\bB}_h^{\ctrl}-\bB_h(\overline{\bB}_h^{\ctrl})\right) \\
&\quad+\alpha_{\rho}J_{\rho}^T\left(J_{\rho}\overline{\bB}_h^{\ctrl}-\rho_h(\overline{\bB}_h^{\ctrl})+\bar{\rho}\right)+\alpha_{\bJ}J_{\bJ}^T\left(J_{\bJ}\overline{\bB}_h^{\ctrl}-\bJ_h(\overline{\bB}_h^{\ctrl})+\bar{\bJ}\right) \\
&\quad+\alpha_{\mathcal{E}}J_{\mathcal{E}}^T\left(J_{\mathcal{E}}\overline{\bB}_h^{\ctrl}-\mathcal{E}_h(\overline{\bB}_h^{\ctrl})+\bar{\mathcal{E}}\right).
\end{aligned}
\end{align}
Solving~\eqref{eq:linear system control em} gives the instantaneous magnetic control $\bB_h^{\ctrl}$.


\begin{remark}
The local linearization adopted here corresponds to a single Gauss--Newton step for the regularized discrete least-squares objective and is one possible strategy for approximately solving the discrete control problem. It reduces the control computation to a positive-definite linear system, avoiding iterative nonlinear optimization. This approximation is particularly suited to the instantaneous setting, where the prediction horizon is one time step and the linearization is recomputed from the updated particle state at each step. Its accuracy nevertheless depends on the departure of the selected control from the expansion point. 
\end{remark}

\begin{remark}\label{remark: parametrization}
Instead of using a piecewise constant control over the spatial grid cells, one may introduce a parametrization to preserve a continuous periodic structure and reduce the number of control variables. Here, as before, $\bF$ denotes either the electric control $\bE^{\ctrl}$ or the magnetic control $\bB^{\ctrl}$. A typical choice is to represent the control using sine and cosine functions. For example, in one spatial dimension,
\begin{equation}\label{def: parametrization}
\bF(x)=\sum_{k=1}^K a_k\sin(kk_0x)+\sum_{k=0}^K b_k\cos(kk_0x), \qquad k_0=\frac{2\pi}{L_x},
\end{equation}
where $\boldsymbol{\theta}=(a_1,\ldots,a_K,b_0,\ldots,b_K)^T$ is the new control variable. The number of control variables is then reduced from the number of spatial grid cells to $2K+1$, depending on the choice of $K$. The linearization, Jacobian computation, and solution of the first-order optimality condition are carried out with respect to $\boldsymbol{\theta}$ instead of $\bF_h$. When the parametrized control is used in the PIC scheme for the particle update, it is first evaluated on the spatial grid according to~\eqref{def: parametrization} and then interpolated to the particle positions. 
This parametrization may be useful when a smooth periodic control is desired. We tested this approach for the two-stream instability using the same numerical setup as in Section~\ref{Sec:numerical}, with the control representation being the only difference. We considered $K=1,3,5,7,9$, corresponding to $3,7,11,15,19$ control variables, respectively. When $K=1$, the control became less effective at later times. Specifically, the electric-field energy began to increase around $t=30$, possibly due to the limited number of control variables. For $K\geq 3$, the control performance was similar to that of the piecewise constant representation. As these numerical results did not show a clear improvement in control performance, we do not further consider this parametrization in this study. 
\end{remark}


\begin{remark}\label{remark: MPC}

In the present formulation, the objective accounts only for the one-step evolution under the current control. To incorporate the dynamics over multiple future time steps, one may instead adopt model predictive control (MPC), in which a sequence of controls is optimized over a finite prediction horizon~\cite{MRRS00,GSS22}. Rather than using the entire control sequence, only the first control is taken as the control for the current time step, and the optimization problem is solved again at the next time step using the updated state. This allows the current control to be chosen based on its effect beyond the next time step. From this perspective, the instantaneous control considered in this work corresponds to a prediction horizon of one time step. A longer prediction horizon, however, increases the computational cost, since multiple forward time steps must be evaluated within each optimization problem.

In our setting, an MPC formulation with a prediction horizon of $N_{\mathrm{h}}$ time steps can be written as an extension of~\eqref{def: instan opt prob}:
\begin{equation}
\begin{aligned}
\left\{\bF^{n+j}\right\}_{j=0}^{N_{\mathrm{h}}-1}
=
\operatorname*{arg\,min}_{\{\bF^{n+j}\}_{j=0}^{N_{\mathrm{h}}-1}}
\quad &
\sum_{j=0}^{N_{\mathrm{h}}-1}
\mathcal{J}^{n+j}
\left(
f\!\left(\bF^n,\ldots,\bF^{n+j}\right),
\bF^{n+j}
\right) \\
\text{s.t.}\quad &
\text{time-discretized version of the controlled system},\\
&
\text{\eqref{eq: controlled vlasov} together with \eqref{eqn: poisson eqn} or \eqref{Gauss's law}--\eqref{Faraday's law}}.
\end{aligned}
\end{equation}
Here, $f(\bF^n,\ldots,\bF^{n+j})$ denotes the predicted state evolution generated by the control sequence $\bF^n,\ldots,\bF^{n+j}$ and used to evaluate the $(n+j)$-th stage cost.
We applied this MPC formulation to the two-stream instability control problem using the same numerical setup as in Section~\ref{Sec:numerical}. We tested prediction horizons $N_{\mathrm{h}}=3,5,7$, with $N_{\mathrm{h}}=1$ corresponding to the instantaneous control. However, extending the prediction horizon did not lead to a substantial change in control effectiveness. In view of these results and the additional computational cost of MPC, we leave further investigation for future work.

\end{remark}

\section{Numerical results}\label{Sec:numerical}

In this section, we present numerical results to demonstrate the performance of the proposed control framework. We first consider the two-stream, bump-on-tail, and Dory--Guest--Harris instabilities in the electrostatic case in Section~\ref{sec: numerical_electrostatic}. We then consider the Weibel instability in the electromagnetic case in Section~\ref{sec: numerical_electromagnetic}. All simulations were performed on either the Hyak Klone cluster or the Doppio server at the University of Washington. We used NVIDIA L40S or RTX 6000 GPUs on Hyak Klone and NVIDIA TITAN X (Pascal) GPUs on Doppio, which is maintained by the Department of Applied Mathematics.

The weighting parameters in the objective functionals~\eqref{obj: electrostatic} and~\eqref{obj: electromag} are chosen separately for each numerical test. We briefly describe the empirical procedure used to select these weights. In our numerical experiments, we observe that the choice of weighting parameters plays an important role in the effectiveness of the resulting control. Using the same optimization procedure with different weighting parameters can lead to different control performance. At the same time, a satisfactory choice of parameters is not unique, and multiple combinations can give comparable control results. Among the combinations that give effective control, we select the one that best reduces the field energy while keeping the distribution close to its initial state.
We select the weighting parameters using the following procedure: We fix the field energy weights at $\gamma_{\bE}=1$ in the electrostatic case and $\gamma_{\bE}=\gamma_{\bB}=1$ in the electromagnetic case. For the regularization parameter $\beta$, we typically begin with values around $0.01$ or $0.001$. A smaller value allows a larger control magnitude, while a larger value more strongly limits the magnitude of the control. For the moment-based weights, we first test
\[
(\alpha_{\rho},\alpha_{\bJ},\alpha_{\mathcal{E}})
=(1,1,1),\quad (1,0,0),\quad (0,1,0),\quad \text{and}\quad (0,0,1),
\]
to examine the combined and individual effects of the three moment terms. Based on the resulting field energy trajectories and the evolution of the distribution function, we then adjust the weights by factors of $10$, increasing or decreasing each weight as needed. Once a suitable range is identified, smaller adjustments are used to further refine the parameter combination. This procedure is repeated until satisfactory instability suppression is obtained.


The Jacobian matrices in~\eqref{electrostatic_jacob} and \eqref{electromag_jacob} can be computed either using automatic differentiation or by explicitly differentiating the discrete computational procedure and propagating the sensitivities through each stage of the algorithm. We refer to the latter as ``hand-coded differentiation''. In our implementation, we primarily use automatic differentiation in PyTorch, as it provides flexibility for future extensions, including more complicated time integrators, despite its higher computational cost. A comparison of the computational cost of the two approaches is provided in Table~\ref{table:tsi_comp_cost}.

\subsection{Electrostatic case}\label{sec: numerical_electrostatic}

For all electrostatic examples, we solve the Poisson equation~\eqref{eqn: poisson eqn} using a Fourier spectral method and determine the external electric field control through the instantaneous optimization problem~\eqref{opt prob verlet}. The control field $\bE_h^{\ctrl,n+1/2}$ is piecewise constant over the spatial grid cells. For the local linearization, we use the control from the previous step as the expansion point,
\[
\overline{\bE}_h^{\ctrl,n+1/2}=\bE_h^{\ctrl,n-1/2}.
\]
For the first step, where no previous control is available, we set $\overline{\bE}_h^{\ctrl,1/2}=\mathbf{1}_h$. Here, $\mathbf{1}_h$ denotes the vector of ones defined on the spatial grid. The remaining simulation parameters vary across the examples and are specified in each section.

\subsubsection{Two-stream instability}\label{sec: numerical_tsi}

We first consider the two-stream instability in one spatial and one velocity dimension (1D1V), with normalized charge $q=1$. The equilibrium distribution is given by
\[
f_{\mathrm{eq}}(v)=\frac{1}{2\sqrt{2\pi}}\left(e^{-(v-2.4)^2/2}+e^{-(v+2.4)^2/2}\right),
\]
and the initial distribution is defined by adding a perturbation in space,
\[
f_0(x,v)=\left(1+0.005\cos(kx)\right)f_{\mathrm{eq}}(v),
\]
where $x\in[0,L_x]$, $L_x=2\pi/k$, and $k=0.2$. We use $N_x=100$ spatial grid points, with $P_c=5\times10^3$ particles per spatial cell, giving a total of $N=5\times10^5$ particles. The initial particle positions and velocities are generated by random sampling from $f_0$. We use a Gaussian kernel for regularization and interpolation, with the kernel parameter chosen as the spatial cell size $\Delta x=L_x/N_x$. The smooth kernel is used because the particle-grid coupling is differentiated when computing the Jacobian with respect to the control variable. The time step is $\Delta t=0.1$, and the simulation is run until $t_{\mathrm{final}}=50$.

The target moments in the objective functional are chosen from the equilibrium distribution:
\[
\bar{\rho}=q\int_{\mathbb{R}}f_{\mathrm{eq}}(v)\,\rd v=1,\qquad
\bar{J}=q\int_{\mathbb{R}}v f_{\mathrm{eq}}(v)\,\rd v=0,\qquad
\bar{\mathcal{E}}=q\int_{\mathbb{R}}v^2 f_{\mathrm{eq}}(v)\,\rd v=6.76.
\]
These constant values are used as the target moments at all spatial grid points. The weighting parameters in the objective functional~\eqref{obj: electrostatic} are set to
\[
\gamma_{\bE}=1,\qquad \beta_{\bE^{\ctrl}}=0.01,\qquad
\alpha_{\rho}=0.1,\qquad \alpha_{\bJ}=0.01,\qquad
\alpha_{\mathcal{E}}=1.
\]

\begin{figure}[htp!]
    \centering
    \includegraphics[width=0.6\linewidth]{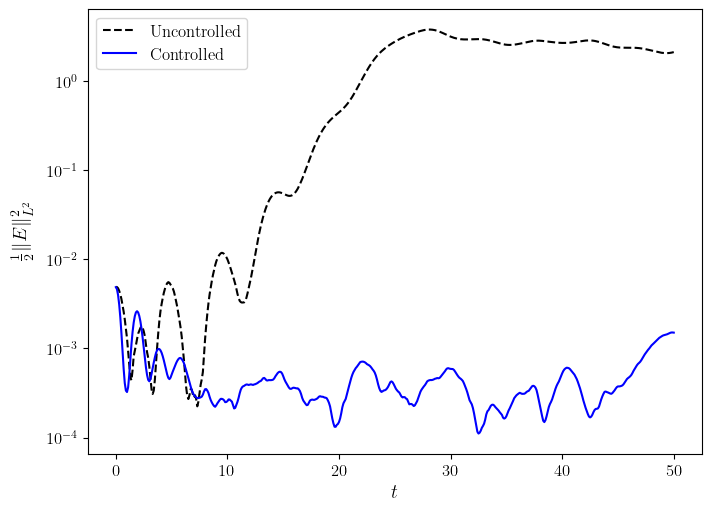}
    \caption{Two-stream instability test. Electric field energy up to $t=50$ for the uncontrolled and controlled cases.}
    \label{fig:TSI_EE}
\end{figure}

Figure~\ref{fig:TSI_EE} compares the electric field energy for the uncontrolled case, shown by the black dashed line, and the controlled case, shown by the blue line. In the uncontrolled case, the electric field energy grows until approximately $t=20$ and then enters a nonlinear plateau, where the linear theory no longer applies. In contrast, the controlled case maintains the electric field energy at a level comparable to or lower than its initial value. This behavior is consistent with our objective design, which includes the reduction of the electric field energy as one of its components.

\begin{figure}[htp!]
    \centering
    \begin{subfigure}[t]{\linewidth}
        \centering
        \includegraphics[width=\linewidth]{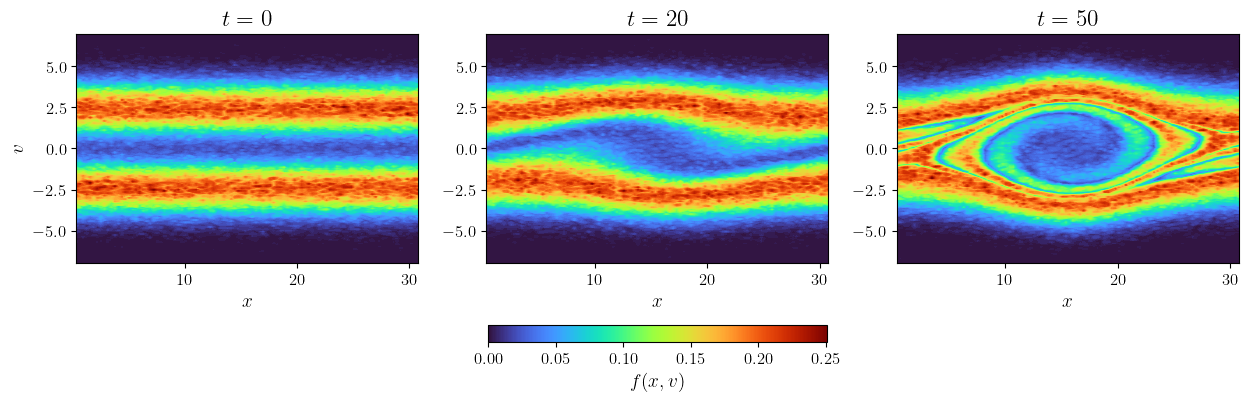}
        \caption{Uncontrolled case}
    \end{subfigure}

    \vspace{0.4cm}

    \begin{subfigure}[t]{\linewidth}
        \centering
        \includegraphics[width=\linewidth]{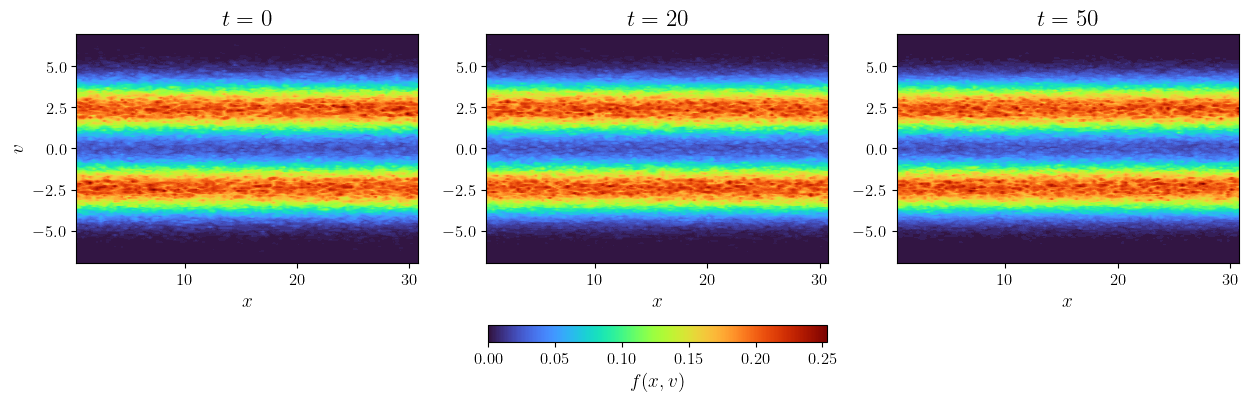}
        \caption{Controlled case}
    \end{subfigure}

    \caption{Two-stream instability test. Time evolution of the phase-space distribution at $t=0$, $20$, and $50$: (a) uncontrolled case and (b) controlled case.}
    \label{fig:TSI_phase}
\end{figure}

The phase-space distribution plots in Figure~\ref{fig:TSI_phase} further show that the control achieves our overall goal of keeping the plasma distribution close to the desired equilibrium while suppressing the instability. The top row shows the uncontrolled case, while the bottom row shows the controlled case. The uncontrolled case develops the typical vortex structure as time evolves, whereas the controlled case remains close to the initial two-stream distribution up to the final time $t_{\mathrm{final}}=50$.
It is worth emphasizing that our control objective does not directly measure the deviation of the distribution function from equilibrium. Instead, it relies only on the electric field energy and three low-order moments, which do not uniquely determine the distribution function. Nevertheless, our control framework successfully preserves the overall structure of the distribution using only this limited information.

Finally, we examine the computational cost of the control procedure using the two-stream instability test as a representative example. Table~\ref{table:tsi_comp_cost} compares the runtime and peak GPU memory usage for two Jacobian computation approaches within the control procedure. With automatic differentiation, the total runtime is 2735.52~s, with 2693.99~s spent in the control procedure. Using the hand-coded Jacobian reduces the control time to 78.10~s and the total runtime to 116.75~s, while the PIC runtime remains comparable between the two cases. For comparison, the corresponding uncontrolled simulation requires 33.67~s. The slightly larger PIC runtime in the controlled simulations includes additional time for data transfer and format conversion. The peak GPU memory is measured over the full simulation, and the automatic-differentiation approach requires approximately 1.6 times the peak memory of the hand-coded approach. All simulations used for this comparison were performed using NVIDIA L40S GPUs on the Hyak Klone cluster.

\begin{table}[t]
    \centering
    \caption{Computational cost of the controlled two-stream instability simulation for two Jacobian computation approaches within the control procedure.}
    \label{table:tsi_comp_cost}
    \begin{tabular}{lrrrr}
        \toprule
        & \multicolumn{3}{c}{\textbf{Runtime (s)}} 
        & \textbf{Peak memory (GB)} \\
        \cmidrule(lr){2-4}
        \textbf{Jacobian method}
        & Total
        & PIC steps
        & Control procedure
        & \\
        \midrule
        Automatic differentiation
        & 2735.52
        & 41.53
        & 2693.99
        & 3.924 \\
        
        Hand-coded differentiation
        & 116.75
        & 38.65
        & 78.10
        & 2.424 \\
        \bottomrule
    \end{tabular}
\end{table}

\subsubsection{Bump-on-tail instability}

We next consider the bump-on-tail instability in 1D1V, with normalized charge $q=1$. The equilibrium distribution is given by
\[
f_{\mathrm{eq}}(v)=\frac{0.9}{\sqrt{2\pi}}e^{-(v+2)^2/2}+\frac{0.1}{\sqrt{0.5\pi}}e^{-2(v-3.5)^2},
\]
and the initial distribution is obtained by adding a perturbation in space to the tail beam,
\[
f_0(x,v)=f_{\mathrm{eq}}(v)+\frac{0.0003}{\sqrt{0.5\pi}}e^{-2(v-3.5)^2}\sin(kx),
\]
where $x\in[0,L_x]$, $L_x=2\pi/k$, and $k=0.2$. We use $N_x=100$, $P_c=5\times10^3$, $N=5\times10^5$, $\Delta t=0.1$, and $t_{\mathrm{final}}=70$. The initial particle positions and velocities are generated by random sampling from $f_0$, and a Gaussian kernel with width $\Delta x=L_x/N_x$ is used.

The target moments are again chosen from the equilibrium distribution:
\[
\bar{\rho}=q\int_{\mathbb{R}}f_{\mathrm{eq}}(v)\,\rd v=1,\qquad
\bar{J}=q\int_{\mathbb{R}}v f_{\mathrm{eq}}(v)\,\rd v=-1.45,\qquad
\bar{\mathcal{E}}=q\int_{\mathbb{R}}v^2f_{\mathrm{eq}}(v)\,\rd v=5.75.
\]
The weighting parameters in the objective functional are set to
\[
\gamma_{\bE}=1,\qquad \beta_{\bE^{\ctrl}}=0.001,\qquad \alpha_{\rho}=0.01,\qquad \alpha_{\bJ}=0.5,\qquad \alpha_{\mathcal{E}}=0.1.
\]

\begin{figure}[htp!]
    \centering
    \includegraphics[width=0.6\linewidth]{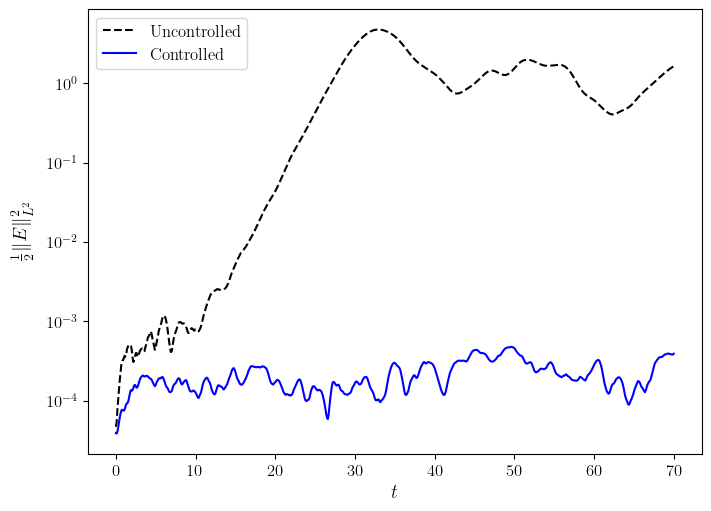}
    \caption{Bump-on-tail instability test. Electric field energy up to $t=70$ for the uncontrolled and controlled cases.}
    \label{fig:BOT_EE}
\end{figure}

\begin{figure}[htp!]
    \centering
    \begin{subfigure}[t]{\linewidth}
        \centering
        \includegraphics[width=\linewidth]{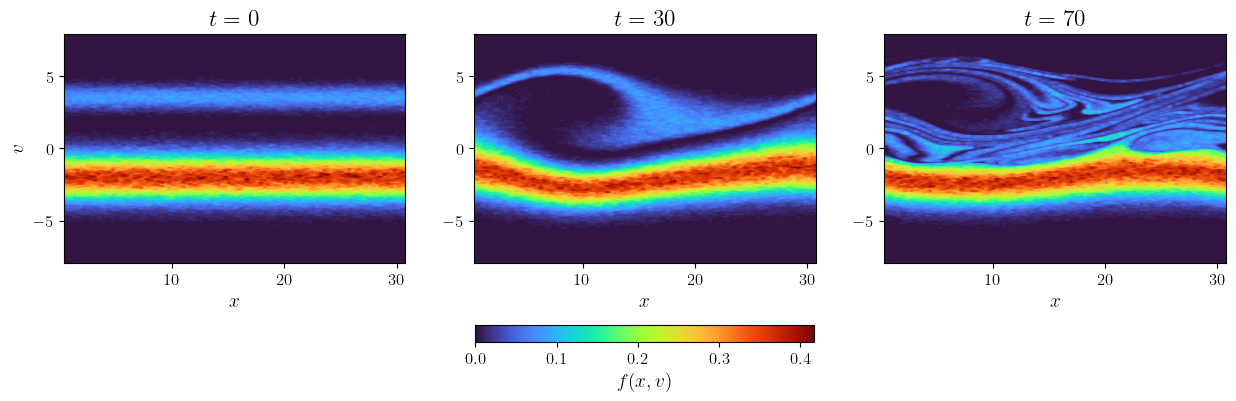}
        \caption{Uncontrolled case}
    \end{subfigure}

    \vspace{0.4cm}

    \begin{subfigure}[t]{\linewidth}
        \centering
        \includegraphics[width=\linewidth]{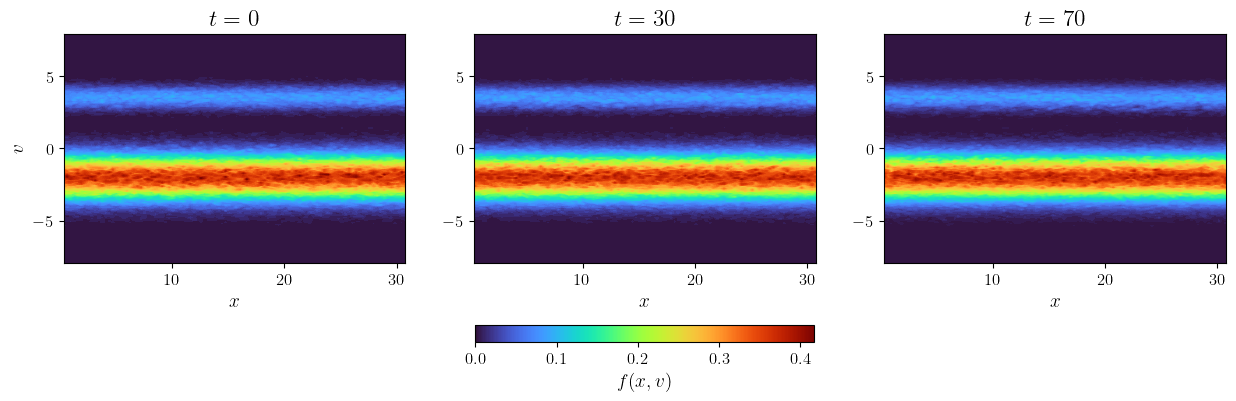}
        \caption{Controlled case}
    \end{subfigure}

    \caption{Bump-on-tail instability test. Time evolution of the phase-space distribution at $t=0$, $30$, and $70$: (a) uncontrolled case and (b) controlled case.}
    \label{fig:BOT_phase}
\end{figure}

Figure~\ref{fig:BOT_EE} shows the electric field energy evolution, while Figure~\ref{fig:BOT_phase} shows the phase-space distributions for the uncontrolled and controlled cases. The bump-on-tail test is similar to the two-stream instability test, but the bulk and tail beams have different velocity widths, and the initial perturbation is applied to the tail beam. As in the two-stream case, the control suppresses the growth of the electric field energy and keeps it near its initial level. The phase-space plots also show that the controlled distribution remains close to its initial shape throughout the simulation.

\subsubsection{Dory--Guest--Harris instability}

We next consider the Dory--Guest--Harris instability \cite{VCS14} in one spatial and two velocity dimensions (1D2V), with normalized charge $q=-1$. The equilibrium distribution is given by
\[
f_{\mathrm{eq}}(v_x,v_y)=\frac{1}{\pi(6!/3)} \left(\frac{v_x^2+v_y^2}{1/3}\right)^6 e^{-(v_x^2+v_y^2)/(1/3)}.
\]
and the initial distribution is obtained by adding a perturbation in space,
\[
f_0(x,v_x,v_y)=f_{\mathrm{eq}}(v_x,v_y)\left[1+0.01\sin(4\theta-kx)\right],
\]
where $\theta=\arctan(v_y/v_x)$. We use $\widetilde{\kappa}=4.65$ as the normalized wavenumber and $v_{\perp0}=\sqrt{2}$, together with a constant background magnetic field $B=0.05$. Here, $v_{\perp0}$ is the radius at which the equilibrium distribution attains its maximum in the $(v_x,v_y)$ plane. The spatial wavenumber is then given by $k=\widetilde{\kappa}B/v_{\perp0}$, and we set $L_x=2\pi/k$ with $x\in[0,L_x]$. The initial particles are generated by deterministic sampling from $f_0$. For this sampling, the domain in $(x,v_x,v_y)$ is divided into $100\times150\times150$ sampling cells, and the center of each cell is used as a particle phase-space location, with $\bv\in[-4,4]^2$. This gives a total of $N=2.25\times10^6$ particles. We use $N_x=48$, $\Delta t=0.1$, and $t_{\mathrm{final}}=400$, with a Gaussian kernel of width $\Delta x=L_x/N_x$.

For the moment based terms, we use the $v_x$ components with the following target values:
\[
\bar{\rho}=q\int_{\mathbb{R}^2}f_{\mathrm{eq}}\,\rd v_x\rd v_y=-1,\qquad
\bar{J}_x=q\int_{\mathbb{R}^2}v_xf_{\mathrm{eq}}\,\rd v_x\rd v_y=0,\qquad
\bar{\mathcal{E}}_{xx}=q\int_{\mathbb{R}^2}v_x^2f_{\mathrm{eq}}\,\rd v_x\rd v_y=-\frac{7}{6}.
\]
The weighting parameters in the objective functional are set to
\[
\gamma_{\bE}=1,\qquad \beta_{\bE^{\ctrl}}=0.003,\qquad \alpha_{\rho}=3,\qquad \alpha_{\bJ}=0,\qquad \alpha_{\mathcal{E}}=10.
\]

\begin{figure}[htp!]
    \centering
    \includegraphics[width=0.6\linewidth]{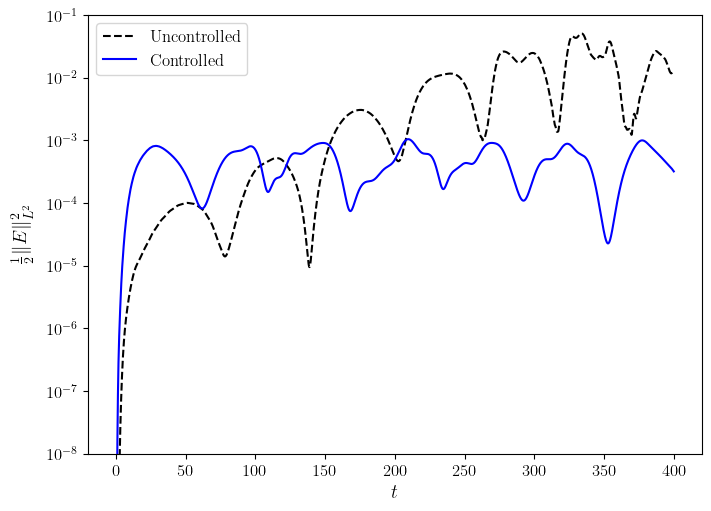}
    \caption{Dory--Guest--Harris instability test. Electric field energy up to $t=400$ for the uncontrolled and controlled cases.}
    \label{fig:DGH_EE}
\end{figure}

\begin{figure}[htp!]
    \centering
    \begin{subfigure}[t]{\linewidth}
        \centering
        \includegraphics[width=\linewidth]{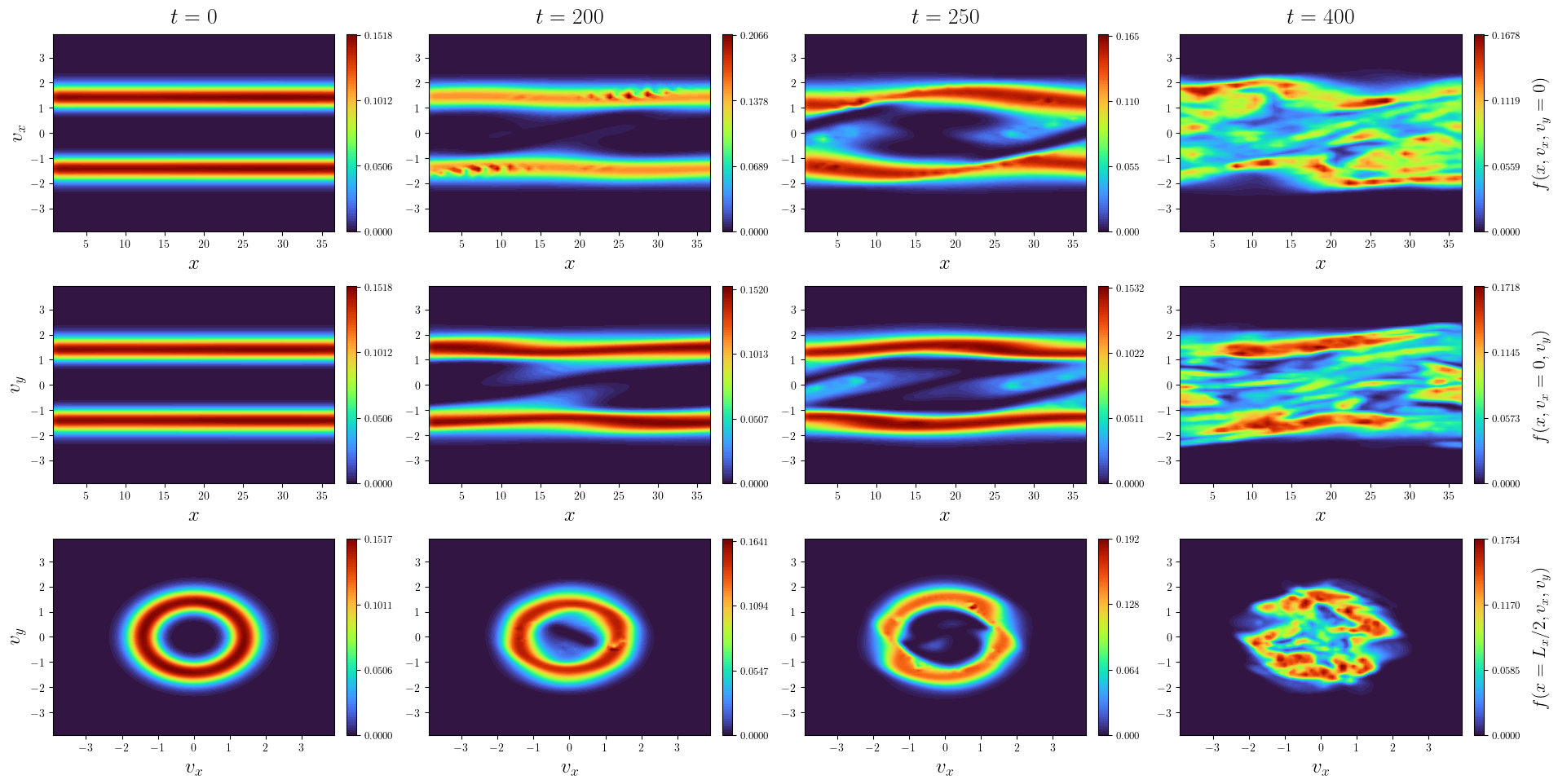}
        \caption{Uncontrolled case}
    \end{subfigure}

    \vspace{0.4cm}

    \begin{subfigure}[t]{\linewidth}
        \centering
        \includegraphics[width=\linewidth]{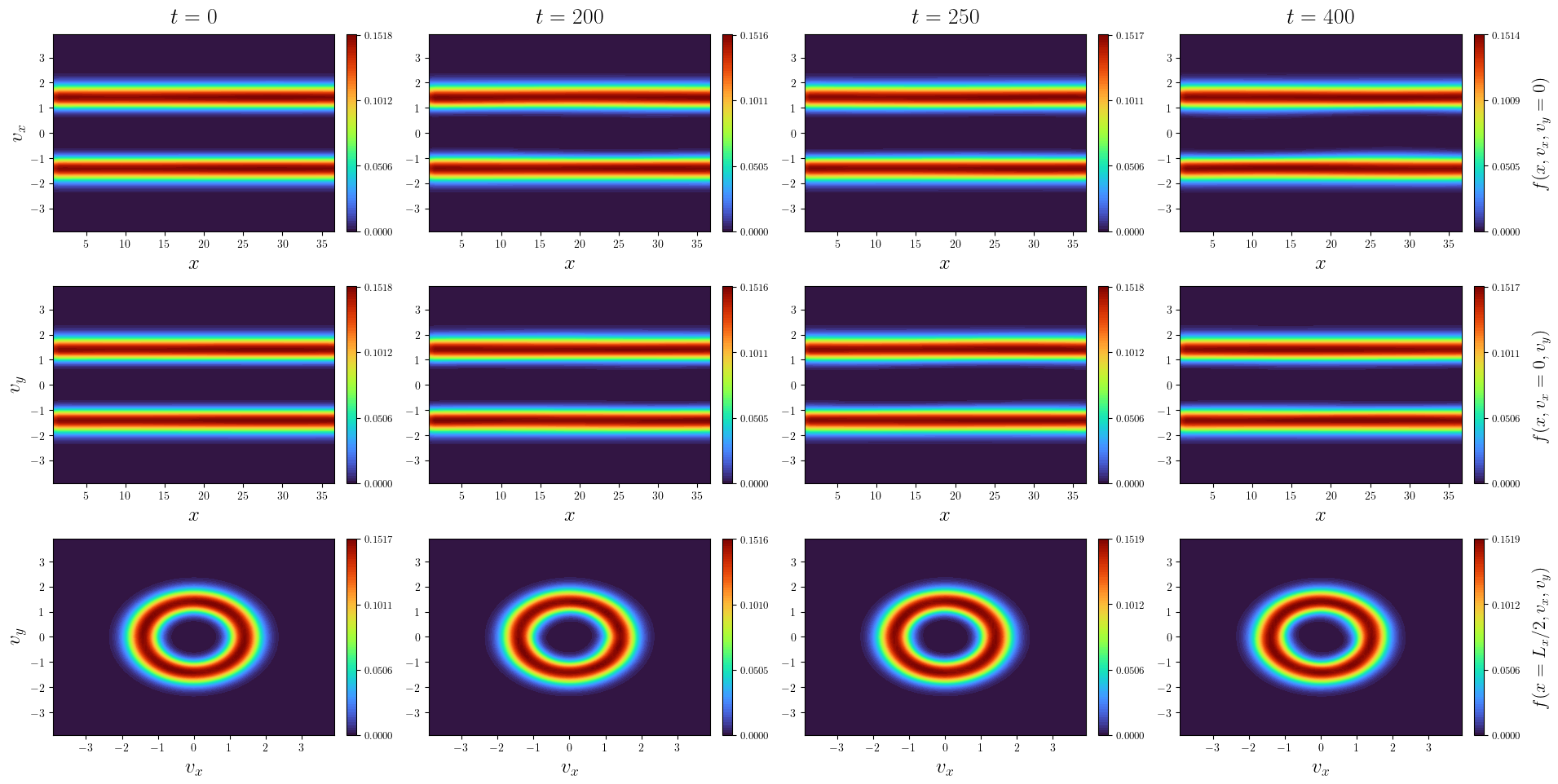}
        \caption{Controlled case}
    \end{subfigure}

    \caption{Dory--Guest--Harris instability test. Time evolution of slices of the distribution function at $t=0$, $200$, $250$, and $400$. From top to bottom, the rows show $f(x,v_x)$ at $v_y=0$, $f(x,v_y)$ at $v_x=0$, and $f(v_x,v_y)$ at $x=L_x/2$, respectively, for (a) the uncontrolled case and (b) the controlled case.}
    \label{fig:DGH_phase}
\end{figure}

Figure~\ref{fig:DGH_EE} shows the electric field energy evolution, while Figure~\ref{fig:DGH_phase} shows the distribution slices in the $(x,v_x)$, $(x,v_y)$, and $(v_x,v_y)$ planes for the uncontrolled and controlled cases. As shown in Figure~\ref{fig:DGH_EE}, the electric field energy in the uncontrolled case continues to grow over a relatively long time interval and reaches a plateau around $t=300$. 
In the controlled case, the electric field energy initially rises above that of the uncontrolled case, but remains at a relatively low level up to $t_{\mathrm{final}}=400$.

The effect of the control is more clearly seen in Figure~\ref{fig:DGH_phase}. In the uncontrolled case, the initial distribution begins to deform around $t=200$ and shows significant deformation by $t=400$. In contrast, the controlled case remains close to the initial distribution in all three distribution slices throughout the simulation. Thus, although the suppression of electric field energy is less pronounced in this test, the control effectively maintains the distribution close to the target equilibrium, which is an important objective of our control strategy. The relatively limited reduction in electric field energy may be related to the more complicated particle dynamics in the presence of a background magnetic field, as well as the competing objectives of reducing field energy and preserving the distribution. A more detailed understanding of these effects is left for future investigation.


\subsection{Electromagnetic case}\label{sec: numerical_electromagnetic}
For the electromagnetic case, we spatially discretize Maxwell's equations using second-order central finite differences on Yee's lattice~\cite{Yee66}. Specifically, the self-consistent electric and magnetic fields are defined on the staggered grids, and the external magnetic control $\bB_h^{\ctrl,n+1/2}$ is defined on the same grid as the self-consistent magnetic field. The control is determined from the instantaneous optimization problem~\eqref{opt prob boris} and is piecewise constant over the spatial grid cells. In the local linearization, the control from the previous time step is used as the expansion point,
\[
\overline{\bB}_h^{\ctrl,n+1/2}=\bB_h^{\ctrl,n-1/2}.
\]
At the first time step, we set $\overline{\bB}_h^{\ctrl,1/2}=\mathbf{1}_h$, where $\mathbf{1}_h$ denotes the vector of ones defined on the grid for the magnetic field.

\subsubsection{Weibel instability}

We consider the Weibel instability in one spatial and two velocity dimensions (1D2V), with normalized charge $q=-1$. The equilibrium distribution is given by
\[
f_{\mathrm{eq}}(v_x,v_y)=\frac{1}{2\pi\beta_{\mathrm{th}}}e^{-v_x^2/\beta_{\mathrm{th}}}
\left(e^{-(v_y-0.3)^2/\beta_{\mathrm{th}}}+e^{-(v_y+0.3)^2/\beta_{\mathrm{th}}}\right),
\]
with $\beta_{\mathrm{th}}=0.01$. We take $f_0(x,v_x,v_y)=f_{\mathrm{eq}}(v_x,v_y)$ and introduce the perturbation through the initial magnetic field. The electric field is initialized as $\bE(0,x)=0$, while the magnetic field is initialized as $\bB(0,x)=(0,0,B_3(0,x))$ with $B_3(0,x)=0.001\sin(kx)$. Here, $x\in[0,L_x]$, $L_x=2\pi/k$, and $k=0.2$. We use $N_x=32$, $P_c=3.2\times10^3$, $N=1.024\times10^5$, $\Delta t=0.1$, and $t_{\mathrm{final}}=120$. The initial particle positions and velocities are generated by random sampling from $f_0$, and a Gaussian kernel with width $\Delta x=L_x/N_x$ is used.


The weighting parameters in the objective functional~\eqref{obj: electromag} are set to
\[
\gamma_{\bE}=1,\qquad \gamma_{\bB}=1,\qquad \beta_{\bB^{\ctrl}}=0.01.
\]
Unlike in the electrostatic cases, we find that the moment penalty is not needed in this example. We also tested the inclusion of moment terms but observed no significant improvement in control performance. 

\begin{figure}[htp!]
    \centering
    \includegraphics[width=0.6\linewidth]{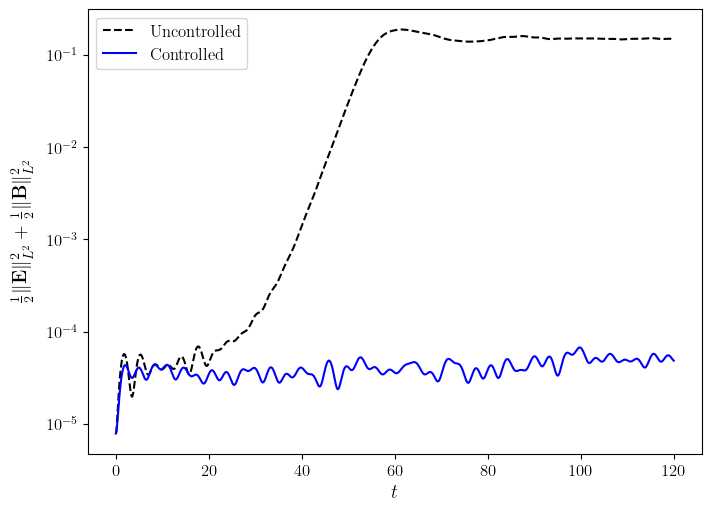}
    \caption{Weibel instability test. Electromagnetic field energy up to $t=120$ for the uncontrolled and controlled cases.}
    \label{fig:weibel_EE}
\end{figure}

Figure~\ref{fig:weibel_EE} shows the electromagnetic field energy, given by the sum of the electric and magnetic field energies. In the controlled case, the control keeps the electromagnetic field energy near its initial level up to $t_{\mathrm{final}}=120$. 

\begin{figure}[htp!]
    \centering
    \begin{subfigure}[t]{0.95\linewidth}
        \centering
        \includegraphics[width=\linewidth]{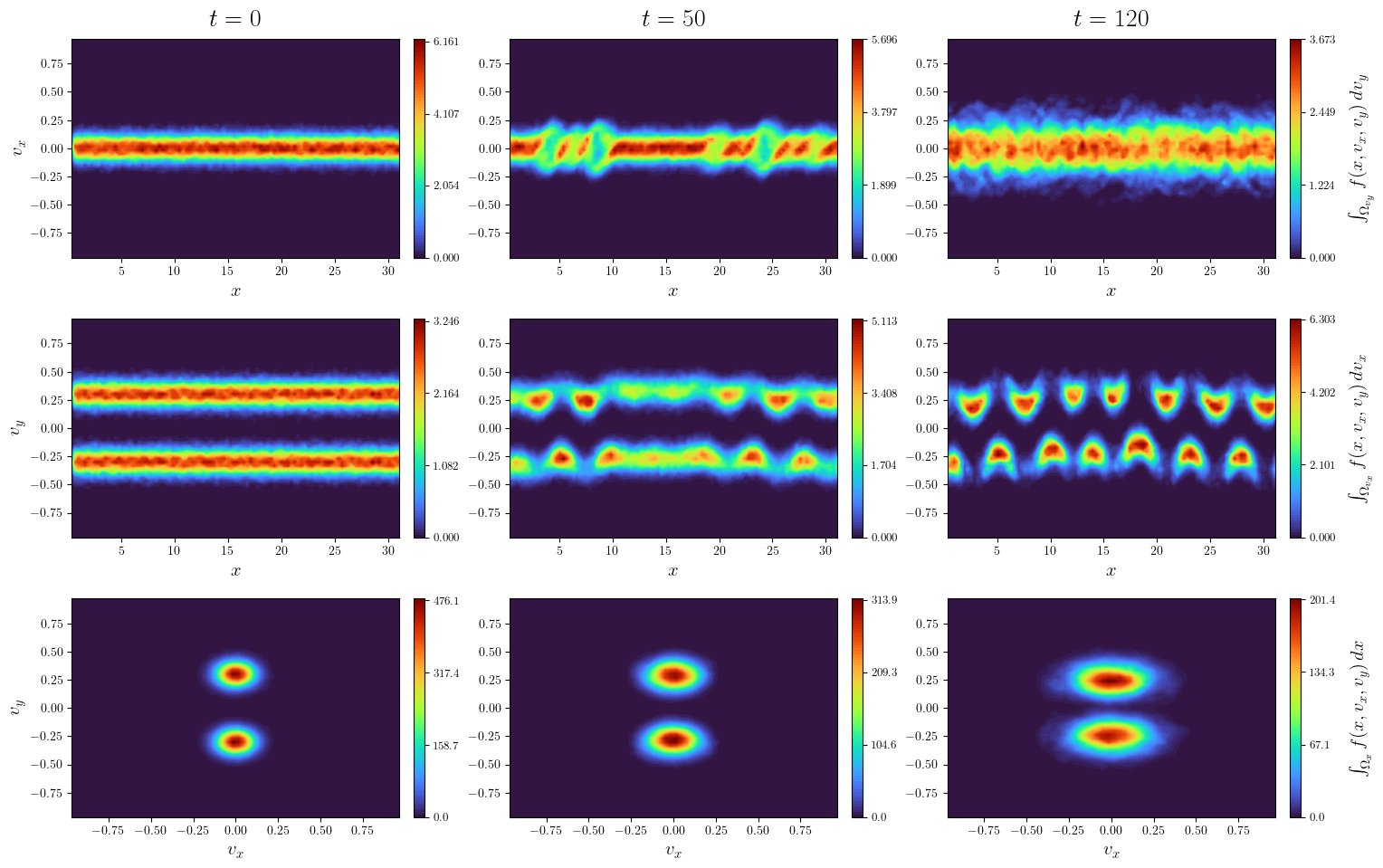}
        \caption{Uncontrolled case}
    \end{subfigure}

    \vspace{0.4cm}

    \begin{subfigure}[t]{\linewidth}
        \centering
        \includegraphics[width=0.95\linewidth]{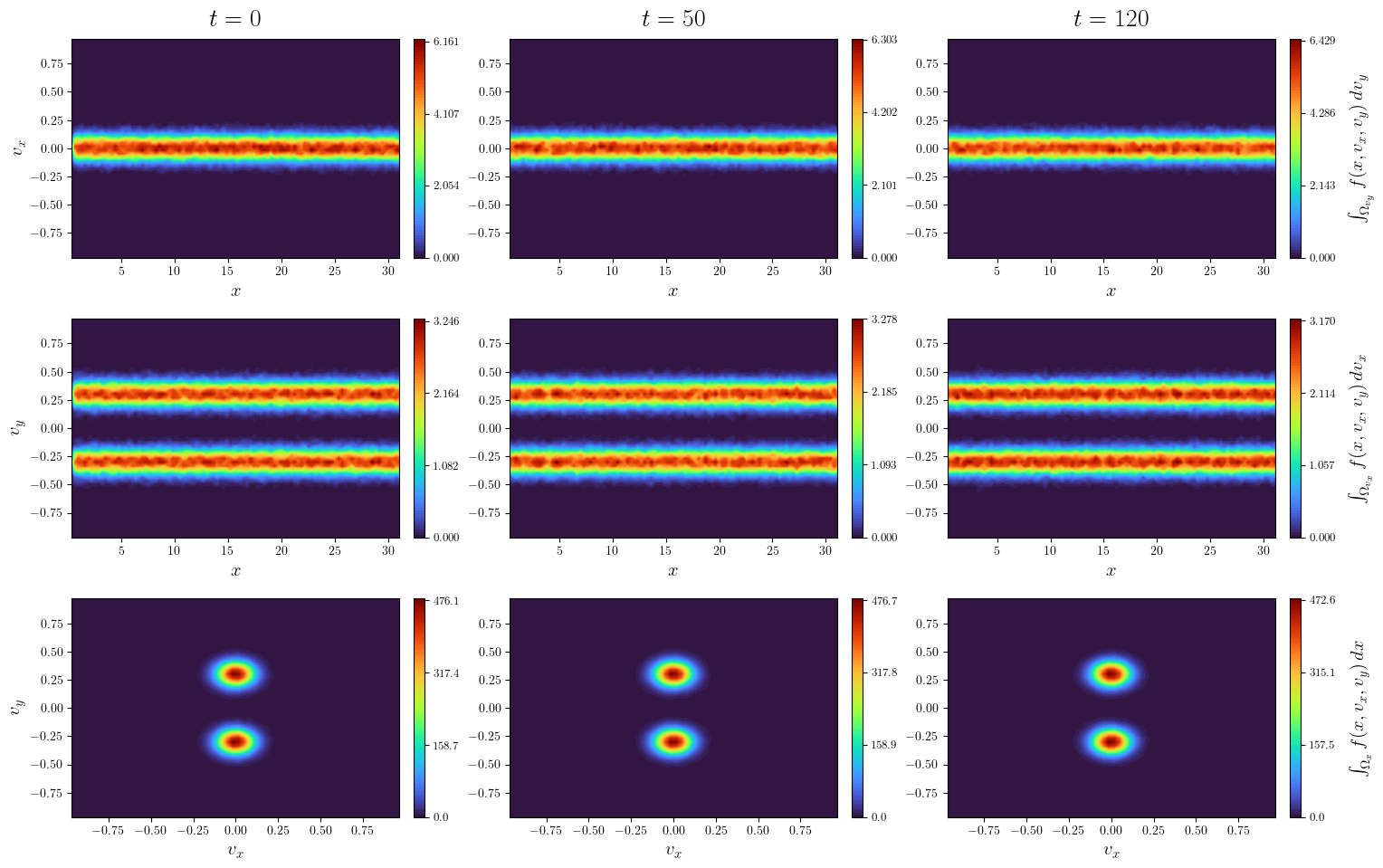}
        \caption{Controlled case}
    \end{subfigure}

   \caption{Weibel instability test. Time evolution of the marginal distributions at $t=0$, $50$, and $120$. From top to bottom, the rows show the $(x,v_x)$, $(x,v_y)$, and $(v_x,v_y)$-marginal distributions, respectively, for (a) the uncontrolled case and (b) the controlled case.}
    \label{fig:weibel_phase}
\end{figure}

Figure~\ref{fig:weibel_phase} compares the marginal distributions for the uncontrolled and controlled cases. In the uncontrolled case, the $(x,v_x)$- and $(x,v_y)$-marginal distributions gradually deviate from their initial shapes, and the $(v_x,v_y)$-marginal distribution exhibits a characteristic spreading of the two velocity beams, with more spreading along $v_x$. In contrast, the controlled case maintains marginal distributions close to their initial shapes at all times shown.

\section{Conclusion}
\label{Sec:conclusion}
In this work, we proposed an instantaneous external field control framework for suppressing kinetic plasma instabilities. The control problem was formulated as a PDE-constrained optimization problem, so that the nonlinear plasma dynamics were directly incorporated into the control design. The control was then determined at each time step, avoiding global-in-time optimization. We used a PIC method as the forward solver and defined the objective function using electromagnetic field energy and low-order moments, together with a regularization term for the control. This allowed the control to be determined without reconstructing the distribution function or comparing it with a target distribution. Through several electrostatic and electromagnetic benchmark problems, we demonstrated the effectiveness of the proposed control strategy in suppressing kinetic plasma instabilities. The numerical results showed that the control suppressed the growth of field energy while maintaining the distribution close to its initial state.

The present study focuses primarily on the formulation and numerical demonstration of the proposed control approach, while a theoretical understanding of the resulting stabilization remains limited. Future work will explore more theoretically grounded control strategies within this framework. Further investigation is also needed to understand how the weighting parameters in the objective functional affect control performance and to identify underlying principles for their selection, rather than relying solely on empirical tuning. The framework can also be extended beyond the current collisionless setting to kinetic plasma models with collisional effects. Another direction is to incorporate structure-preserving particle methods, including energy-conserving approaches~\cite{RH25,YHR26}, to preserve important structural properties of the plasma dynamics under control.

\section*{Acknowledgement}

This work was partially supported by DOE grant DE-SC0023164, NSF grants DMS-2409858 and IIS-2433957, and AFOSR grants FA9550-24-1-0254 and FA9550-26-1-B161. Both authors thank Lee Ricketson for the helpful discussion of plasma particle simulations.


\bibliographystyle{abbrv}
\bibliography{hu_bibtex}

\end{document}